\documentclass[a4paper,fleqn,usenatbib]{mn2e}
\usepackage{amssymb}
\usepackage{textcomp}
\usepackage{graphicx}
\usepackage{txfonts}
\usepackage{epstopdf} 
\usepackage{xcolor}     

\begin{document}

\title[2021 PH27 as active asteroid]{Is 2021 PH27 an active asteroid with a meteor shower detectable on Venus?
\thanks{}}
\author[Carbognani, Tanga, Bernardi]{Albino Carbognani,$^{1}$ \thanks{E-mail: albino.carbognani@inaf.it}
Paolo Tanga,$^{2}$, Fabrizio Bernardi,$^{3}$\\
$^{1}$INAF - Osservatorio di Astrofisica e Scienza dello Spazio, Via Gobetti 93/3, 40129 Bologna, Italy\\
$^{2}$Universit$\acute{e}$  Nice C$\hat{o}$te d'Azur, Observatoire de la C$\hat{o}$te d'Azur, Laboratoire Lagrange UMR7293 CNRS, Nice, France\\
$^{3}$SpaceDyS, 56023 Navacchio di Cascina, Pisa, Italy
}


\date{Received ; Accepted}

\maketitle

\label{firstpage}

\begin{abstract}
The recently discovered near-Earth asteroid 2021 PH27 has the shortest orbital period of all known asteroids. It cannot be excluded that 2021 PH27 is also an active asteroid, as (3200) Phaethon. We intend to estimate the consequences of this hypothesis, although testing is difficult with ground-based observations during perihelion passages, due to low solar elongation. Assuming a surface activity similar to that of Phaethon, an increase in brightness of about 1.4 mag can be estimated. Since it is an asteroid with a MOID of $0.014660 \pm 0.000034$ AU with Venus, 2021 PH27 could be the equivalent of Phaethon for the Earth and be the progenitor body of a venusian meteor shower. A good opportunity to observe the hypothetical fireballs in Venus's atmosphere will take place on the days around Jun 07, 2023, when Venus will pass at the minimum distance from the nominal orbit of 2021 PH27. Another favorable date will be Jul 05, 2026. Finally, on Mar 28, 2022 the asteroid will also be at the maximum Sun elongation of about $52.3^{\circ}$ and at the aphelion of its orbit, the most favorable configuration to characterize it from the physical point of view with photometric, polarimetric and spectroscopic observations.
\end{abstract}

\begin{keywords}
minor planets, asteroids: individual: 2021 PH27 
\end{keywords}

\section{Introduction}
\label{sec:introduction}
The electronic circular MPEC 2021-Q41\footnote{https://www.minorplanetcenter.net/mpec/K21/K21Q41.html} published by the Minor Planet Center (MPC) on Aug 21, 2021, announced to the astronomical community the discovery of the near-Earth 2021 PH27 which holds the record for the asteroid with the shortest known orbital period: only about 114.5 days to complete a full orbit around the Sun.\\
This asteroid belongs to the class of Atiras, minor bodies whose orbits are completely inside the Earth's one: they have an aphelion smaller than the Earth's perihelion of 0.983 AU. The Atiras - named after the first object of this class, (163693) Atira discovered in 2003 - are the smallest group of near-Earth asteroids, compared to the most populated - and easy to discover - groups: Atens (2096 bodies), Apollos (13706) and Amors (11285). 
Currently 50 Atira are known\footnote{https://minorplanetcenter.net/mpc/summary}. The discovery of this class of asteroids is difficult because they are observable at low angular distances from the Sun, only immediately after sunset or just before the rise of our star, when the sky is still clear and the use of a telescope becomes problematic \citep{ye2020}.\\

2021 PH27 was discovered by Scott Sheppard of the Carnegie Institution of Science during the twilight of Aug 13, 2021 using the Dark Energy Camera (DECam)
connected to the V\'ictor M. Blanco Telescope of 4 metres aperture at F/2.94 of the Cerro Tololo Observatory, La Serena (Chile): by exploiting ``dead time'' of the instrument during twilight. Subsequent observations made by David Tholen (University of Hawaii) and Marco Micheli (ESA), allowed to confirm the presence of this elusive member of the solar system. 2021 PH27 has an asteroidal orbit (semimajor axis = 0.462 AU, eccentricity = 0.712, inclination = $31.9^{\circ}$), with an asteroid-like Tisserand parameter with respect to Jupiter: $T_J = 11.6$\footnote{JPL Small-Body Database Browser, https://ssd.jpl.nasa.gov/}. From MPC astrometric data the absolute magnitude (hereafter mag) is +17.7 with a diameter $D_n$ value between 0.5 and 1.5 km if the geometric albedo ranges from 0.5 and 0.05: in this letter we assume a diameter of about 1 km. The taxonomic class, shape and rotation period are unknown.

\section{A comparison between Phaethon and 2021 PH27}
\label{sec:venusian_phaethon}
From the MPC website 44 astrometric observations have been made from 13 Aug to 8 Sep 2021, plus two further recovered positions of 16 Jul 2017 which significantly lengthened the observed time span. A total of 46 position measurements are therefore available with 0.46 arcsec rms using NEODyS system. The MOID (Minimum Orbit Intersection Distance) with the Earth is 0.225 AU - so not a particularly low value for a near-Earth - while the MOID with Venus is much lower. Taking into account the orbital uncertainties we estimated a value of $V_{MOID} = 0.014660 \pm 0.000034$ AU, with the Venus-MOID flyby which occurs when the planet has the true anomaly equal to about $82.68^{\circ}$. This low MOID value would imply, in case the asteroid was active, that it could generate a meteoroid stream swept by Venus. In other terms, 2021 PH27 could be ``the Phaethon of Venus''. In what follows we will explore the consequences of our hypothesis, implying that the asteroid should be carefully observed during the next favorable elongation in the period Mar-Apr 2022.\\

\subsection{Phaethon}
\label{sec:phaethon}
The active asteroid (3200) Phaethon - equivalent diameter of about 5.5 km \citep{taylor2019} - has a MOID with Earth of about 0.0194 AU and is considered the main meteoroids source in the Geminid meteor shower: together with the asteroids 2005 UD (MOID = 0.077 AU) and 1999 YC (MOID = 0.247 AU) they form the Phaethon-Geminid complex, which probably originated from the fragmentation of a common ancestor \citep{jewitt2010}. The discovery of Phaethon's activity was not easy. At its maximum elongation from the Sun, Phaethon shows no sign of gas or dust emission. Only the observations made with the NASA's STEREO-A probe in 2009 with Phaethon near the perihelion showed photometric anomalies associated with a surface activity because the brightness increased of about 2 mag as the phase angle increased \citep{li2013}: a trend exactly opposite to what is reported about normal asteroids \citep{carbognani2019}. \\
Therefore, for Phaethon, the time interval of about 24 days required to reach a Sun distance comparable to that of Venus, starting from perihelion, is sufficient to cancel any form of remotely observable activity. Indeed, deep observations made in Dec 2017 when Phaethon was only 0.07 AU from Earth and about 1 AU from the Sun showed no signs of activity, i.e. no coma or the presence of nearby fragments larger than about 15 metres in diameter \citep{ye2021}.\\

Explaining the physical phenomenon that makes Phaethon active is more complex. The asteroid rotates in about 3.6 hours, i.e. it is far from the classic spin-barrier value of about 2.2 hours: so it is not an activity triggered by rotational disintegration as for (6478) Gault \citep{carbognani2021}, also because it would not be limited to perihelion alone. The activity in conjunction with perihelion, suggests the sublimation of volatile materials that cannot be cometary ices because, given the high temperatures experienced by the surface (about 1000 K near perihelion), they would have already been sublimated and - as before - it would not be limited to perihelion. According to this picture, recent near-IR observations found no evidence of hydration on Phaethon surface \citep{takir2020}. The most plausible actual physical mechanism appears to be the sodium sublimation from the asteroid's rock matrix, assisted by thermal fracturing of the surface. This mechanism would explain why the Geminid meteoroids are so low in sodium \citep{masiero2021}.\\
We do not know if the sodium emission encompasses the entire surface or only certain areas. What we know is that Phaethon surface is not uniform. Radar observations made by Arecibo showed the presence of a depression more than 1 km across (probably an impact crater) near the equator and a 600-m dark region near one of the poles \citep{taylor2019}. The photometric observations do not show a variation of the color indices, so probably mineralogy is homogeneous, while a variation of the linear polarization between 47 and 50 degrees of the phase angle has been observed, probably due to a change of the regolith properties \citep{borisov2018}.

\subsection{2021 PH27 activity estimate}
\label{sec:activity}
In this section we will assume that Phaethon and 2021 PH27 are asteroids of the same type. \\
Considering that the perihelion of 2021 PH27 is inferior to Phaethon's perihelion, it is reasonable to suppose that its activity could be maintained by the same physical processes, and that it shows similar properties. In particular, the activity could be restricted to a range of distances near perihelion, as in the discovery images there is not a coma. Using the observed Phaethon activity as a guide, we can make a rough estimate of the expected increase in brightness for 2021 PH27.\\
In absence of constraints about the surface homogeneity of 2021 PH27, we assume for simplicity that the activity for both asteroids is evenly distributed over the surface. In this case, with the asteroids both at perihelion and considering the diameters, we can expect that the 2021 PH27 dust emission is reduced by a factor equivalent to the size ratio $\left( 1/5.5\right)^2 \approx 0.03$ with respect to Phaethon. So, if the cross section of the added dust for Phaethon in 2009 and 2012 was $C \approx 100 ~\hbox{km}^{2}$ \citep{li2013}, for 2021 PH27 it can be expected to be $C\approx 100\cdot 0.03\approx 3 ~\hbox{km}^{2}$. The geometrical cross section of the nucleus is $C_n = \pi \left( D_{n}/2 \right)^{2}\approx 0.8 ~\hbox{km}^{2}$, therefore the increasing factor of the brightness flux with respect to the bare core will be $C/C_n \approx 3.7 $, hence an increase equal to $\Delta m = 2.5 Log\left( 3.7 \right)\approx 1.4$ mag can be expected. This brightness increase is not negligible but the difficulty of an observation near perihelion is considerable, even for a probe as STEREO-A, because 2021 PH27 is an object about 3.4 mag weaker than Phaethon. If the decrease in activity with distance is the same as for Phaethon, it will not persist at large solar elongations: 2021 PH27 takes 36 days to pass from perihelion to about 0.72 AU (the semi-major axis of the orbit of Venus), a value 12 days higher than Phaethon.\\
Still in the hypothesis that 2021 PH27 is an active asteroid similar to Phaethon, we can estimate the average density of meteoroids present along the orbit and therefore the associated meteoroid stream density. The number of meteoroids per unit of volume present in space, $N_m$, will be proportional to the surface of the asteroid $R^2$ and its orbital frequency $f$ (the greater the number of perihelion passes, the larger the amount of ejected meteoroids). Also, it will be inversely proportional to the orbital perimeter $l$ (the longer the orbit, the more diluted the meteoroids will be: the density variation along the orbit is not considered) and to the orbital frequency $F$ of the intersecting planet (each Venus passage in proximity of the MOID slightly reduces the population). We can write:

\begin{equation}
N_m \propto R^2 \left(\frac{1}{l}\right)f\left(\frac{1}{F}\right).
\label{APR}
\end{equation}

Using Eq. (\ref{APR}) and computing the ratio of Phaethon's number of meteoroids with 2021 PH27 we find ($P$ is the index for Phaethon, $V$ for Venus and $E$ for Earth):

\begin{equation}
\frac{N_P}{N_m} = {\left(\frac{R_P}{R}\right)}^2 \left(\frac{l}{l_P}\right)\left(\frac{f_P}{f}\right)\left(\frac{F_V}{F_E}\right)\approx 30\cdot 0.3\cdot 0.2\cdot 1.6\approx 3.
\label{APR2}
\end{equation}

Therefore, assuming the ages of streams are comparable, we can expect the mean number of meteoroids per unit volume associated with the hypothetical stream of 2021 PH27 to be about 1/3 of Geminids. From this rough estimate, the meteoroid stream hypothetically associated with 2021 PH27 appears to be not negligible.

\section{Observation of 2021 PH27}
\label{sec:observe}
In section~\ref{sec:activity} we have mentioned that, if 2021 PH27 {\sl has a similar activity at perihelion as Phaethon}, there are good reasons to doubt that it can be detected when at the maximum elongations from the Sun, i.e. when it is observable from Earth in the evening or morning sky in conditions similar to Venus. In any case, the times around maximum elongations are the ideal moment for its physical characterization, both for photometry (determination of the rotation period, shape and the surface colors), polarimetry and spectroscopy (to highlight surface inhomogeneity if any). \\
The next favorable opportunity to observe 2021 PH27 will be on Mar 28, 2022 at about 10 UT when the asteroid will be at the maximum geocentric solar elongation of about $52.3^{\circ}$: a few hours later, at 17:40 UT, it will be in aphelion of its orbit. The asteroid will be observable in the morning sky in the months of Mar and Apr, 2022. As its declination will be negative ( $-13^{\circ}$/$-22^{\circ}$), observers in the southern hemisphere will be privileged. Unfortunately, the phase angle will change very little in the two months (from $91^{\circ}.5$ to $86^{\circ}.7$) or even anticipating observations to Feb ($100^{\circ}$) so the characterization of the scattering properties by photometry and polarimetry will remain difficult. Other favorable conditions will then occur in future opportunities on Jul 15, 2022, with an evening elongation of $44.7^{\circ}$, and in the morning of Mar 11, 2023 (elongation $48.9^{\circ}$). For a more complete list see Table~\ref{t00}.

\begin{table}
\centering
\caption{Summary of the maximum elongations of 2021 PH27 from the Sun in the period 2022-2026.}
\label{t00}
\begin{tabular}{lc}
\hline
Data & Sun elongation ($^{\circ}$)  \\

\hline
28-Mar-2022 & 52.3  \\
15-Jul-2022 & 44.7  \\
11-Mar-2023 & 48.9  \\
28-Jun-2023 & 49.8  \\
22-Feb-2024 & 42.2  \\
10-Jun-2024 & 50.3  \\
04-Feb-2025 & 34.2  \\
30-May-2025 & 37.7  \\
16-Apr-2026 & 36.6  \\
09-Ago-2026 & 35.1  \\

\hline
\end{tabular}
\end{table}

\subsection{Visibility of a hypothetical meteor shower}
\label{sec:meteor_shower}
A further method to highlight the activity of 2021 PH27 would be to try observing meteors in Venus's atmosphere. In fact, the MOID of 2021 PH27 in respect to Venus's orbit (0.0146 AU) is lower than the same MOID of Phaethon in respect to Earth's orbit (0.0194 AU), so we are in the geometric conditions to have a meteor shower in the atmosphere of Venus. The use of the atmosphere of Venus as a meteoroid detector was suggested long ago (see \cite{ryabova2019} for a review) and was recently also proposed by \cite{zhang2021} regarding comet C/2021 A1 (Leonard).\\

From the orbits geometry of Venus and 2021 PH27, the MOID is only reached near the descending node of the asteroid orbit, so there is only one point of the orbit of Venus in which the observation of the meteors in the planet's atmosphere is theoretically possible. To maximize the observation probabilities, it is reasonable to observe in the periods in which Venus and 2021 PH27 are also at the minimum distance, so as to intercept the most massive part of the hypothetical meteoroid stream, but any time Venus is at the MOID with the orbit of 2021 PH27 may be the right chance.\\

On Mar 17, 2022 at about $7$ UT\footnote{JPL Small-Body Database Browser, https://ssd.jpl.nasa.gov/}, 2021 PH27 will be at the minimum nominal distance of about 0.058 AU from Venus. The planet will be clearly visible in the morning sky from Earth, at an angular distance of about $46.5^{\circ}$ from the Sun, with disk phase 0.48 and apparent diameter of 25.7 arcsec. However, the highest probability of observing the hypothetical meteor shower is with Venus exactly at the MOID (true anomaly $82.68^{\circ}$), i.e. two days earlier, on Mar 15, 2022 at 06 UT with a distance Venus-Earth $\sim$ 0.638 AU, disk phase 0.46 and apparent diameter of 26.4 arcsec.

\begin{table}
\centering
\caption{Cytherocentric apparent ecliptic latitude and longitude of Sun, Earth and true radiant meteor shower computed for March 15, 2022 at 06 UT.}
\label{t01}
\begin{tabular}{lcc}
\hline
Target & Long ($^{\circ}$) & Lat ($^{\circ}$) \\

\hline
Sun             & 034.1  & -02.3 \\
Earth           & 127.9  & -02.6 \\
Meteor radiant  & 335.8  & +20.4 \\

\hline
\end{tabular}
\end{table}

The cytherocentric velocity vector of the meteoroids reaching Venus is given by the difference between the heliocentric velocity vectors of the meteoroids and Venus. It provides the direction of the true radiant as seen by a venusian observer. The vector difference between the positions of the Sun and the Earth with Venus gives the cytherocentic directions in which the Sun and the Earth are visible, see Table~\ref{t01}. For practical reasons we express all directions in terms of ecliptic coordinates. \\
From JPL's HORIZONS system we retrieved the heliocentric positions and velocities of Venus and 2021 PH27 when they respectively reach the MOID between the two orbits, thus assuming that the orbital position and velocity of 2021 PH27 correspond to the average for the hypothetical meteoroids that it ejects. In the case of 2021 PH27 it occurs on March 10, 2022 at 8 UT (asteroid true anomaly $165.42^{\circ}$). \\ 

\begin{figure}
\hspace*{-0.8cm}\includegraphics[width=1.1\hsize]{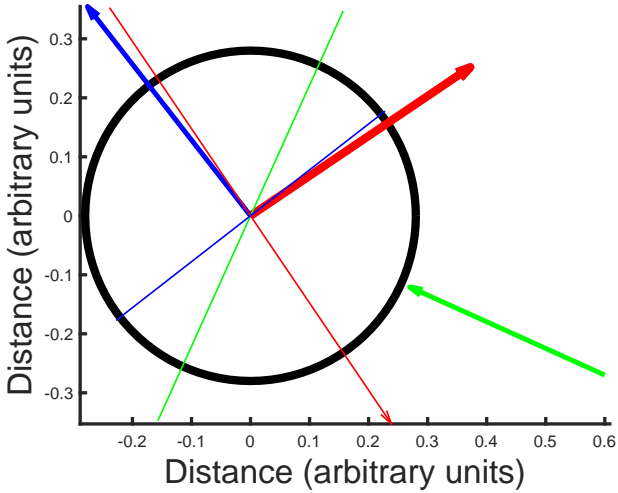}
\caption{The projected geometry on the ecliptic plane of the Sun, radiant and Earth as seen from Venus (black circle) on Mar 15, 2022, during the planet's flyby with the MOID of 2021 PH27. The planar geometry is representative of the 3D geometry due to low inclinations. The X and Y axes are distances in arbitrary units. The red arrow indicates the direction of the Sun, the blue one that of the Earth and the green one is the speed of the meteoroids. Thin lines with the same color indicate the respective terminators (the terminator relative to the Sun has an arrow in Venus's motion direction). Unfortunately, the sub-radiant point is located in the diurnal hemisphere of Venus, not visible from Earth.}
\label{fig:geovenus1}       
\end{figure}

\begin{figure}
\hspace*{-0.8cm}\includegraphics[width=1.1\hsize]{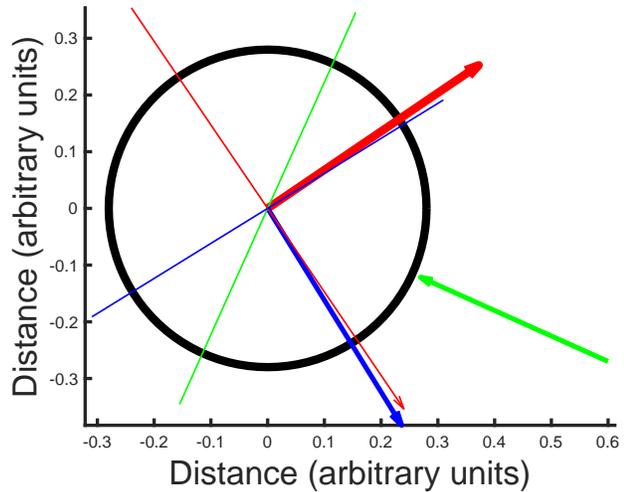}
\caption{The geometry on the ecliptic plane of the Sun, radiant and Earth as seen from Venus (black circle) on Jun 07, 2023. Fireballs may be observable because they partly fall into Venus's dark side visible from Earth. Same color code and axes units as in Fig.~\ref{fig:geovenus1}.}
\label{fig:geovenus2}       
\end{figure}

\begin{figure}
\hspace*{-0.8cm}\includegraphics[width=1.1\hsize]{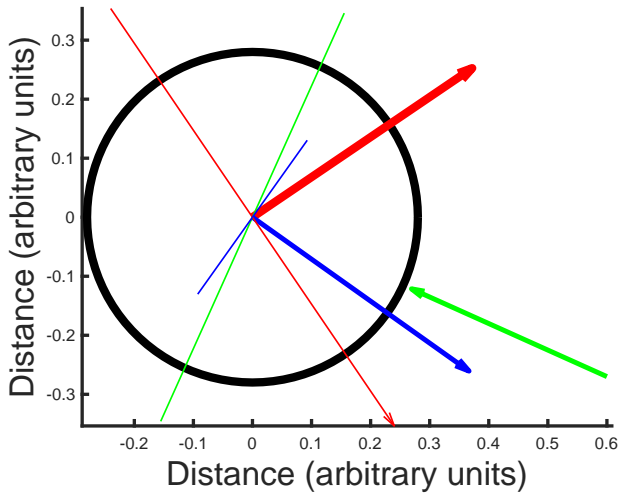}
\caption{The geometry on the ecliptic plane of the Sun, radiant and Earth as seen from Venus (black circle) on Jul 05, 2026. Fireballs may be observable from Earth because they partly fall into Venus's dark side visible from our planet. Same color code and axes units as in Fig.~\ref{fig:geovenus1}.}
\label{fig:geovenus3}       
\end{figure}

\noindent The computations show that the angular distance between the Sun and the meteor radiant will be about $i=61.5^{\circ}$, therefore the fraction of the Venus disk illuminated by the Sun and visible from the radiant is $k=(1+\cos(i))/2\approx 0.74$ and the meteors will be visible only if they fall into the remaining fraction of the disk, $k'= 1-0.74 = 0.26$. \\
On Mar 15, 2022 the angular distance between the Sun and the Earth seen from Venus will be about $93.7^{\circ}$, see Fig.~\ref{fig:geovenus1}, so it follows that the sub-radiant point - where the maximum probability to see fireballs is - falls in the diurnal hemisphere of Venus not visible from Earth. This encounter geometry between Venus and the hypothetical meteoroids associated with 2021 PH27 is always the same, regardless of the position of the Earth along its orbit. It follows that in the morning elongations of Venus no impacts from 2021 PH27 meteoroids will be visible from Earth, while the evening elongations are more favorable. \\
We searched for other dates when Venus is at the MOID with the orbit of 2021 PH27 and the results are listed in Table~\ref{t02}. Excluding cases in which the elongation of Venus from the Sun is too low, the illuminated fraction of the planet from Earth too high or when the angle between the direction of the Earth and that of the radiant is very large, there are two suitable dates for the observation of any fireballs in Venus's dark side: June 7, 2023 (see Fig.~\ref{fig:geovenus2}) and July 5, 2026 (see Fig.~\ref{fig:geovenus3}). In both cases it is an evening elongation, with the planet comfortably observable in the early evening.

\begin{table}
\centering
\caption{Visibility summary from Earth of possible 2021 PH27 fireballs in Venus's atmosphere in the period 2022-2026. The column marked SE is the Venus-Sun elongation in degree, IF is the illuminated fraction of Venus's disk visible from Earth, RE is the angle meteors radiant-Earth in degrees, $\Delta$ is the distance between Venus and 2021 PH27 in AU on the date while FV is for fireballs visibility in Venus's dark side of the disk. In bold type are the most favorable dates for observing the possible meteors in Venus's atmosphere.}
\label{t02}
\begin{tabular}{lccccc}
\hline
Data   & SE & IF & RE & $\Delta$ & FV\\

\hline
15-Mar-2022  &  46.5  &  0.47  &  147.6 & 0.0631 & No\\
25-Oct-2022  &  01.2  &  1.00  &  60.0  & 0.0165 & No\\
\bf{07-Jun-2023}  &  \bf{45.4}  &  \bf{0.48}  &  \bf{40.5}  & \bf{0.3960} & \bf{Yes}\\
18-Jan-2024  &  33.5  &  0.83  & 106.8  & 0.1181 & No\\
30-Aug-2024  &  23.6  &  0.91  &  32.0  & 0.1909 & No\\
11-Apr-2025  &  28.0  &  0.12  & 156.4  & 0.2746 & No\\
22-Nov-2025  &  11.0  &  0.98  &  74.8  & 0.3709 & No\\
\bf{05-Jul-2026}  &  \bf{41.7}  &  \bf{0.67}  &  \bf{24.6}  & \bf{0.4810} & \bf{Yes}\\

\hline
\end{tabular}
\end{table}

\subsection{Fireball flux estimate and apparent brightness}
\label{sec:meteor_mag}
In this final section we make an estimate of the flux and the apparent mag that bright fireballs associated with 2021 PH27 could have. The relative speed of the meteoroids with respect to the planet is about $V_{mr} \approx 23.2$ km/s so taking into account Venus's escape velocity $V_e = 10.1$ km/s, the impact velocity in the atmosphere is $V_{i}=\sqrt{{V_{mr}}^2+{V_e}^2} \approx \sqrt{23.2^2+10.1^2} \approx 25.3 $ km/s. This speed is significantly slower than Geminids (about 35 km/s), thus resulting in less bright meteors for a same meteoroid mass. Considering that the zenithal hourly rate for Geminids is $ZHR\sim$100 $~\hbox{h}^{-1}$, that this quantity is proportional to the entry velocity and the meteoroids number density and remembering Eq. (\ref{APR2}), for the hypothetical meteoroid stream associated with 2021 PH27 we have: 
$ZHR\approx 100\cdot \left(25/35\right) \cdot \left(1/3\right)\approx 24~\hbox{h}^{-1}$, an intensity comparable to our Orionids. The relation between the $ZHR$ and the shower flux $F(<M)$, where $M$ is the absolute mag and $F$ is the flux measured in meteors $ ~\hbox{h}^{-1} ~\hbox{km}^{-2}$, was given by \citep{ehlert2020}:

\begin{equation}
F\left(<+6.5\right) = \frac{ZHR\left(13.1r-16.5\right)\left(r-1.3\right)^{0.748}}{37200~\hbox{km}^{2}},
\label{flux}
\end{equation}

\noindent where $r$ is the population index of the shower. Using the Geminids' value, $r\approx 2.6$ \citep{ehlert2020}, with $ZHR\approx 24~\hbox{h}^{-1}$ we find an estimated flux 
$F(<+6.5)\approx 0.014 ~\hbox{meteors} ~\hbox{h}^{-1} ~\hbox{km}^{-2}$.\\
In the evening of Jun 07, 2023 Venus will be at a distance of about 0.682 AU from Earth, so a meteor in the planet atmosphere will appear $\sim 30$ magnitudes fainter than a terrestrial meteor of equal brightness seen at 100 km of altitude. Under these conditions, from Earth, we can only hope to observe the brightest fireballs, such as those of absolute mag $V\sim$ -12/-15. \\
From the flux values of $F(<+6.5)$ above, we can get the flux of the brightest fireballs with absolute mag less than -15 by the distribution \citep{ehlert2020}:

\begin{equation}
F\left(< M_1\right) = F\left(< M_2\right)\left[10^{(M_2-M_1)/2.5}\right]^{1-\alpha}.
\label{flux_bright}
\end{equation}

In Eq. (\ref{flux_bright}) $\alpha$ is the mass index of the shower. If we assume $M_1 = -15$, $M_2=+6.5$ and $\alpha\approx 1.95$ (same as Geminids), we get $F(< -15)\approx 10^{-9} F(<+6.5)$. Considering the impact section of Venus (a circular area projected on a plane) and the shaded fraction exposed to the radiant, $k'\approx 0.26$, one can expect to see a fireball of mag less than -15 about every 300 hours. This number is clearly not very encouraging for observations, but given all the several assumptions and the absence of constraints on the real flux, it can just be considered as a rough indication maybe off by a large factor.\\ 
In the case of July 5, 2026, the geocentric distance of Venus will be about 1 AU and the observation of fireballs even more difficult. In both cases, the diffused light from the illuminated fraction of the planet disk will disturb unless shielded in the optical path. Considering the uncertainties involved in the extension of the meteoroid stream, it is recommended to start observing a few days before and after the flyby of Venus with the MOID.\\
A much more promising alternative to ground observations is provided by space probes. The Japanese Akatsuki probe (JAXA)\footnote{https://akatsuki.isas.jaxa.jp/en/mission/} is currently operating around Venus on an elliptical orbit with a period of 10 days. The distance between the spacecraft and Venus ranging from 1,000 km to 370,000 km. The Lightning and airglow camera (LAC) of Akatsuki can detect lightning discharge in Venusian atmosphere and would be perfect for detecting bright meteors too \citep{zhang2021}. Unfortunately the spacecraft operates for about 30 minutes each orbital revolution, a severe limitation. This makes ground observations even more important.\\

\section{Summary and conclusions}
The purpose of this letter is to draw observers' attention to the recently discovered Atira-type asteroid 2021 PH27. We do not know whether 2021 PH27 is an active asteroid but based on what we know about Phaethon, this is possible. Its direct observation near perihelion (where an activity is most likely to occur) is difficult. A geocentric maximum elongation from the Sun of 2021 PH27 will be reached on March 28, 2022 at $52.3^{\circ}$, and will be a first good opportunity for its full physical characterization, preferably by southern observers. Other favorable conditions will then occur on Jul 15, 2022, with an evening elongation of $44.7^{\circ}$, and in the morning of Mar 11, 2023 (elongation $48.9^{\circ}$). 

An anomalous brightness increase induced by the ejected dust should be detectable in the asteroid photometry. This would strongly help to better constrain the density of the hypothetical meteoroid swarm. In fact, by simply assuming an activity similar to Phaethon, it appears very difficult to observe from the ground the bright fireballs associated with any meteor shower that should impact Venus's atmosphere. \\
The optimum chances for the next few years will be around Jun 07, 2023 and Jul 05, 2026 during favorable Venus-MOID flyby. In both cases the planet will not be too far from disk dichotomy and a part of the meteoroids stream - if any - could impact the nocturnal hemisphere partially visible from Earth. Unfortunately, only very bright fireballs can be observed from the ground over several hours of observation, while current and future spacecrafts around the planet could have better chances from their vantage positions.\\

\section*{acknowledgements}
Many thanks to SpaceDyS for use of the code for MOID computation and to Giovanni Valsecchi for useful advice. Many thanks to the referee Jeremie Vaubaillon who, with his comments, has increased the quality of the manuscript. For this paper JPL's HORIZONS system (on-line solar system data and ephemeris computation service), was also used. 

\section*{Data Availability}
The data underlying this article will be shared on reasonable request to the corresponding author.


\bsp
\label{lastpage}
\end{document}